\documentclass[prc,aps,nofootinbib,twocolumn]{revtex4}
\usepackage{epsfig}
\usepackage{graphicx}
\usepackage{amssymb}
\usepackage{color}
\usepackage{upgreek}
\usepackage{mathtools}
\usepackage[dvipsnames]{xcolor}
\usepackage{xfrac}  
\usepackage{nicefrac}

\usepackage{braket}

\begin{document}

\title{Entanglement in selected Binary Tree States: Dicke/Total spin states, particle number projected BCS states }

\author{Denis Lacroix } \email{lacroix@ijclab.in2p3.fr}
\affiliation{Universit\'e Paris-Saclay, CNRS/IN2P3, IJCLab, 91405 Orsay, France}

\date{\today}
\begin{abstract}
Binary Tree States (BTS) are states whose decomposition on a quantum register basis formed by a set of qubits can be made sequentially. 
Such states sometimes appear naturally in many-body systems treated in Fock space when a global symmetry is imposed, like the total spin or particle number symmetries. 
Examples are the Dicke states, the eigenstates of the total spin for a set of particles having individual spin $1/2$, or states obtained by projecting 
a BCS states onto particle number, also called projected BCS in small superfluid systems. 
Starting from a BTS state described on the set of $n$ qubits or orbitals, the entanglement entropy of any subset of $ k$ qubits is analyzed.  Specifically, a practical method is developed to access 
the $k$ qubits/particles von Neumann entanglement entropy of the subsystem of interest. Properties of these entropies are discussed, including scaling properties, upper bounds, or how these entropies correlate with fluctuations.  Illustrations are given for the Dicke state and the projected BCS states.      
\end{abstract}


\maketitle

\section{Introduction}

With the progress in quantum computing \cite{McC16,Cao19,McA20,Bau21,Bha21,End21,Ayr23} or in novel techniques to treat many-body systems, like tensor networks \cite{Cir21}, 
the understanding of entanglement in interacting systems found in physics or chemistry is attracting more attention.  
In recent years, several attempts have been made to better characterize quantum entanglement in nuclear physics \cite{Rob21,Joh23,Paz23,Bul23,Bul23-b,Per23}. Among the current discussions on the subject, one can mention the important role of spontaneous symmetry breaking \cite{Fab21, Fab22} that might be connected to a quantum phase transition or might characterize correlations in many-body systems \cite{Dus04,Vid04a,Vid04b,Vid04c,Dit19}, or the "volume/area law" nature of entanglement \cite{Gu23}. quantifying entanglement might also be very useful in practice for improving 
the convergence of many-body theories, as was firstly shown in quantum chemistry \cite{Ste16,Ste19} and more recently analyzed in nuclear physics \cite{Mom23,Tic23}. The possible measure of entanglement through the study of fluctuations in nuclear reactions has also been 
discussed very recently \cite{Li24}.

Most of the studies so far aim to study the one particle or, eventually, two particles entanglement using a variety of tools, like the von-Neumann reduced entropy, Renyi entropy, and mutual information \cite{Nie00,Ami08,Hor09,Laf16,Sri24}... In this work, I point out that, due to some specificities of some of the states used in many-body systems, 
 one can access more generally their $k$-particles entropy. Specifically, I use the fact that symmetries affect entanglement by creating block structures in 
 reduced-density matrices. As an illustration, in Ref.  \cite{Lac22}, we have empirically shown that 
the entanglement entropy of $k$ interacting neutrinos with $k>1$  acquires a rather simple scaling property compared to the one neutrino entropy.
Such a scaling can be partially understood from the permutation invariance of the problem. This finding was one of the motivations for the present work.     

In this work, I consider a certain class of states relevant to the symmetry problem in fields like quantum chemistry when total spin is conserved 
or in nuclear physics when treating superfluity using projected quasi-particle states. Following Ref. \cite{Kha23,Dut23}, where a quantum computer algorithm is proposed to build such states, they will be called here  
generically Binary-Tree-State (BTS). BTS states include the Dicke states \cite{Dic54}, the total spin eigenstates in systems formed of particles with spin \cite{Mes62,Low69}, and, as we will see, the so-called 
projected BCS states (BCS states) \cite{Rin80,Bla86,Del01,Bri05}. The entanglement properties of the Dicke state are rather well-known and will serve as a guide for other BTS states below. Still, a study of the general entanglement properties of projected quasi-particle states has yet to be made.    

The present discussion is relevant for many-body systems but can have some interest in the context of quantum computing. For this reason, 
my starting point will be to consider a set of two-level systems, with levels labeled $0$ and $1$,  that will be called hereafter generically qubits. When working with a set of particles with spins $1/2$, each of the two levels will represent the two possible spins of the particle. When considering a many-body problem, each two-level is assigned to a single particle orbital treated in Fock space, and its occupation is assigned to the level $1$ while its vacancy is assigned to $0$.

The article is organized as follows. First, BTS states are introduced, and a method to evaluate the entanglement of any bipartition of the system 
is developed. Then, two illustrative cases are considered: the fully permutation invariant Dicke state and the projected BCS state. Special attention is paid to the entanglement entropy's scaling properties and upper limits.           

\section{Binary Tree States}

I consider here a set of $n$ qubits labelled by $i=1, \cdots, n$, where each qubit can access two states $\{ | 0_i \rangle, |1_i \rangle \}$, and I denote below $\{ X_i, Y_i, Z_i\}$
the corresponding Pauli matrices. Connection with many-body systems described in Fock space can be made through the Jordan-Wigner transformation 
\cite{Jor28,Lie61} giving a mapping 
between the occupation or not of single-orbital with the occupation of the $|1_i\rangle$ and $|0_i \rangle$ state, respectively. 
A state $| \Psi \rangle$ can be decomposed in the qubit register as:
\begin{eqnarray}
| \Psi \rangle &=& \sum_{i=1, \cdots , n; \delta_i =0,1} \Psi_{\delta_{n}, \cdots , \delta_{0}} | \delta_{n}, \cdots , \delta_1 \rangle,  \label{eq:gen}  
\end{eqnarray}
where the little-endian convention is used to order qubits indices.  The states $\{ | \delta_{n}, \cdots , \delta_1 \rangle \}$ will be called 
qubit register basis below.  

In this work, I follow Ref. \cite{Kha23} and  firstly consider a specific class of states that are written as:
\begin{eqnarray}
| K, n \rangle &=& \frac{1}{\sqrt{I^n_K}} [B^+_n (x_1, \cdots x_n)]^K | 0 \rangle^{\otimes n},  \label{eq:agp}
\end{eqnarray} 
for $K=1,\cdots,n$ and   
where $| 0 \rangle^{\otimes n} \equiv |0_{n}, \cdots, 0_{1} \rangle$. The operator $B^+_n$ 
is parametrized in terms of a set of complex numbers $\{ x_1, \cdots, x_n \}$ as  
\begin{eqnarray}
B^+_n  (x_1, \cdots x_n)&=& \sum_{i=1}^n x_i \sigma^+_i , 
\end{eqnarray}
where $\sigma^+_i = (X_i + i Y_i)/2$ is the standard rising operator, with $|1_i \rangle = \sigma^+_i |0_i \rangle$, and 
$\sigma^+_i |1_i \rangle = 0$.  In the following, I will write simply $B^+_n$ when no confusion is possible. Finally,     
$I^n_K$ is a normalization factor insuring $\langle K, n | K, n \rangle = 1$. 

Using the fact that $\left[ \sigma^+_i \right]^2 =0 $ for all $i$, a direct development of $[B^+_n (x_1, \cdots x_n)]^K$ shows 
that the state $| K, n \rangle$  corresponds to a weighted sum of states of the register basis all 
having a Hamming weight equal to $K$, i.e. all states correspond to a binary string that verifies $\sum_i \delta_i =K$. 

States defined by Eq. (\ref{eq:gen}) are of special interest for quantum chemistry or physics. In the specific case where the $\{ x_i \}_{i=1, \cdots, n}$ are the same for all $i$, these states 
identify with so-called Dicke states \cite{Dic54}. An important property 
of these states is that they are invariant with respect to the permutation of any couple of qubits indices. When qubit states are mapped to particles with spins, i.e. 
$\{ | 0_i \rangle , | 1_i \rangle \} \equiv \{ | \downarrow \rangle_i , | \uparrow \rangle \}_i $, the Dicke states given by Eq. (\ref{eq:agp}) are nothing but the eigenstates of the total spin operators 
${\bf S}^2$ having the maximal allowed value $S=n/2$ for $n$ particles. The explicit correspondence between the Dicke states and the angular momentum states 
denoted by $|S,M \rangle$ with $-S\le M \le S$ can be made using the fact  
\begin{eqnarray}
M = K - n/2. 
\end{eqnarray}    
Those states are fully symmetric, i.e., they correspond to states having a single line in their Young tableau and are directly connected to the permutation symmetry group \cite{Mes62}. 

Another physical situation where these states appear for non-equal $\{x_i\}$ is 
for small superfluid systems \cite{Del01}. To see this, one can start  from a non-normalized BCS state written in Fock space as \cite{Bri05}: 
\begin{eqnarray}
| \Phi (\{ x_i \}) \rangle &=&  \prod_{i=1,n} [ 1 + x_i a^\dagger_i a^\dagger_{\bar i} ]| - \rangle, \label{eq:bcs}  
\end{eqnarray}     
where $(a^\dagger_i , a^\dagger_{\bar i} )$ are creation operators of pairs of time-reversed single-particle states, and $| - \rangle$ is the Fock space vacuum.  
To make connection with the state (\ref{eq:agp}), one can introduce the pair creation operator $P^\dagger_i = a^\dagger_i a^\dagger_{\bar i}$, and make the direct SU(2)
mapping of the pair occupation with the occupations of the states $|1_i \rangle $. With this, we have the correspondence $P^\dagger_i \leftrightarrow \sigma^+_i$. This mapping has been used first in Ref. \cite{Kha21} and subsequently in \cite{Rui22,Rui23,Lac23,Rui23b} 
in the context of quantum computing to reduce the qubits number 
at the price of restricting the description to seniority zero many-body states.  Starting from this BCS state mapped on qubits, the state $| K, n \rangle$ is obtained by normalizing the state (\ref{eq:bcs}) after projecting it onto the pair number, or equivalently Hamming weight, equal to $K$.  These states have been extensively employed in many-body systems, especially in nuclear physics \cite{Rin80,Bla86,Bri05,San08,San09,Lac10,Hup11}. 

I focus the discussion here on their entanglement properties. 
 Noteworthy, different techniques have been proposed recently to obtain
these states on a quantum computer, some using indirect measurement techniques \cite{Lac20,Rui22,Rui23,Lac23,Rui23b}, and more recently, direct methods (see  \cite{Kha23} and Refs. therein). In particular, the direct method uses the fact that the state defined by Eq. (\ref{eq:agp}) has a binary tree structure so that each qubit can be sequentially 
introduced one after the other.

 \subsection{k-qubit entanglement entropy: definition and exact results}
 
 I am interested in the entanglement property of a subset of $k$ qubits. Specifically, I will assume that the 
 $n$ qubits separate into two subsystems $\bar A = \{i=1, \cdots , n-k \}$ and $A = \{i=n-k+1, \cdots , n\} $ and  the objective is 
 to access the entanglement property of the subsystem $A$. 
 
 Note that we do not lose any generality in the discussion below by selecting states in $A$ with labels greater than $n-k$. Indeed, one can take 
 an arbitrary set of indices $\{ i_1, \cdots , i_k \}$ and reorganize the indices such 
 that $\{i_1, \cdots , i_k\} \rightarrow \{ i=n-k+1  , \cdots, n\}$. Given a partition of the qubit register $(A, \bar A)$, I introduce the two operators:
 \begin{eqnarray}
B^+_A &=& \sum_{i \in A} x_i \sigma^+_i, ~ B^+_{\bar A} = \sum_{i \in \bar A} x_i \sigma^+_i. \label{eq:bab}
\end{eqnarray} 
I also introduce the two subsets of normalized states:
\begin{eqnarray}
| M, k \rangle_A &=& \frac{1}{\sqrt{I^k_M (A)}} [B^+_A ]^M | 0 \rangle_A,  \label{eq:agp-a}  \\
| J, n-k \rangle_{\bar A} &=& \frac{1}{\sqrt{I^{n-k}_J ({\bar A})}} [B^+_{\bar A}]^J | 0 \rangle_{\bar A},  \label{eq:agp-bara} 
\end{eqnarray} 
with the two conditions $0 \le M \le k$ and $0 \le J \le n-k$.

Starting from a state $| K, n \rangle$ defined in the full qubits register, the reduced von Neumann entropy of the subsystem $A$ is 
defined as
\begin{eqnarray}
S_A &=& - {\rm Tr} \left[ D_A \log_2 D_A \right],\label{eq:entropy-a}
\end{eqnarray} 
with $D_A$, the reduced density given by:
\begin{eqnarray}
D_A &=& {\rm Tr}_{\bar A} \left[ D_{K, n} \right]={\rm Tr}_{1,\cdots, n-k}  \left[ D_{K, n} \right].  \label{eq:da}  
\end{eqnarray} 
$D_{K,n}$ denotes the total density matrix, $D_{K, n} = |K, n \rangle \langle K, n |$.
Since the total state is a pure state, we also have the property $S_{\bar A} = S_A$ \cite{Nie00}.  

\begin{figure}[t] 
\includegraphics[width=\linewidth]{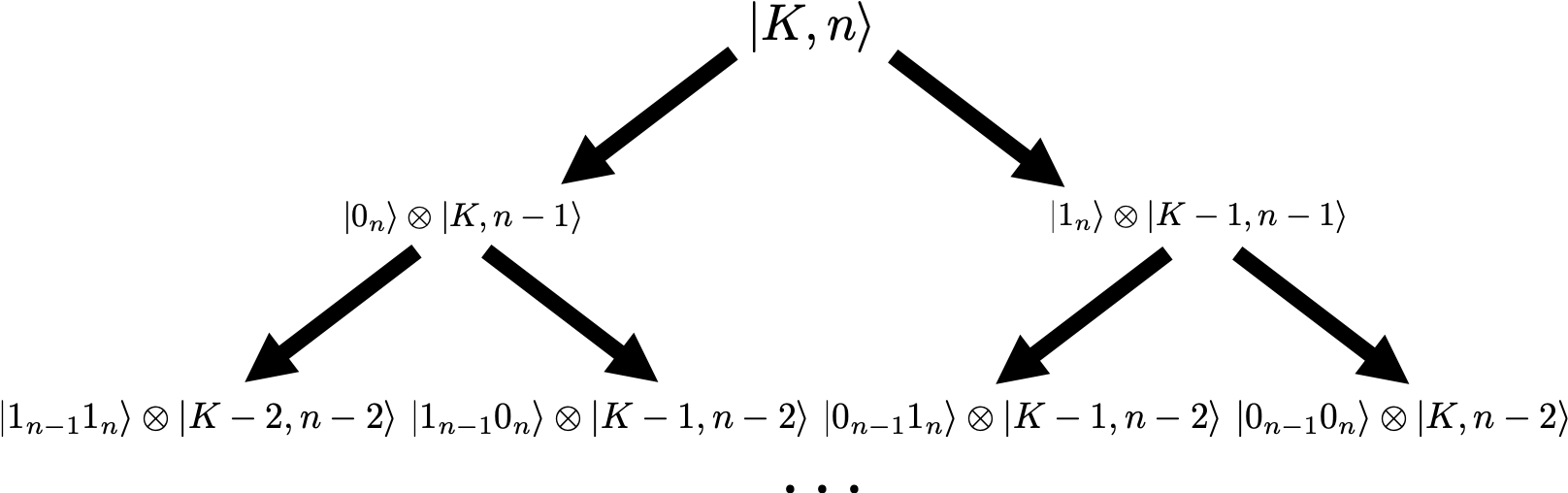} 
    \caption{Schematic view of the binary tree decomposition of the state $| K, n \rangle$ where the first qubit labelled by $n$ is isolated from other qubits 
    following Eq. (\ref{eq:decstate}). The same procedure is then iterated with the qubit labeled by $n-1$ and so on and so forth. }
    \label{fig1:bts}
\end{figure}

In the following, I will use subsystem $A$ with increasing qubits numbers $k$ and sometimes denote simply 
its density and entropy $D_k$ and $S_k$, respectively. The size of this reduced density matrix is $2^k$ and therefore 
increases exponentially with $k$. For small values of $k$, a possible brute-force technique to obtain $S_k$ is to take advantage 
of the BTS structure of the state (\ref{eq:agp}). As an illustration of this structure, let us consider a set of $n$ qubits. If we first focus on $S_1$ where $A$ corresponds to the last qubits $n$. 
By using $B^+_n = B^+_{n-1}(\bar A)  + x_n  \sigma^+_n$, one can easily show that we have:
\begin{eqnarray}
&&| K, n \rangle=  \sqrt{\frac{I^{n-1}_K(\bar A)}{I^n_K}} |K, n-1 \rangle_{\bar A} \otimes |0_n \rangle \nonumber \\
&& ~~~+ x_n K \sqrt{\frac{I^{n-1}_{K-1}(\bar A)}{I^n_K}} |K-1, n-1 \rangle_{\bar A} \otimes |1_n \rangle, \label{eq:decstate}
\label{rec:state}
\end{eqnarray}
leading to the reduced density: 
\begin{eqnarray}
D_1 &=& \lambda_0 |0_n \rangle \langle 0_n | +  \nonumber 
\lambda_1 |1_n \rangle \langle 1_n |,
\end{eqnarray}
with 
\begin{eqnarray}
\lambda_0 &=&  \frac{I^{n-1}_K(\bar A)}{I^n_K}, ~\lambda_1 = K^2 |x_n|^2 \frac{I^{n-1}_{K-1}(\bar A)}{I^n_K}, \nonumber 
\end{eqnarray}
from which the 1-qubit entropy can be computed. One can show that $\lambda_0 + \lambda_1 = 1$. This property
could be directly seen from expression (\ref{eq:decstate}) using that $| K, n \rangle$ is a normalized state.       
$S_2$ can also be obtained by removing similarly the qubit $n-1$ from the $\bar A$ subspace, and so on.   The binary tree decomposition 
of the state is schematically represented in Fig. \ref{fig1:bts}.

This iterative procedure becomes rapidly cumbersome with an exponential increase of terms. It also hides the fact that, in both subspaces 
$A$ and $\bar A$, the states are invariant with respect to the exchange of the ordering of the qubits selected during the state decomposition along the tree. This property is inherited from the permutation invariance of the state $| K, n\rangle$.  

An alternative method to obtain a compact expression of the reduced density is to use simply the fact that 
$B^+_n = B^+_A + B^+_{\bar A}$, leading to:
\begin{eqnarray}
\left[ B^+_n \right]^K &=& \sum_{l=0}^{K} C^l_K \left[ B^+_A \right]^l \left[ B^+_{\bar A} \right]^{K-l}. \label{eq:devB}
\end{eqnarray}  
Assuming that $A$ contains $k$ qubits, some terms in the sum eventually cancel out. Specifically, the term with:
\begin{eqnarray}
l > k ~{\rm or}~ K-l > n-k. \label{eq:constraint}
\end{eqnarray}
This can be interpreted as the fact that not more than $k$ (resp. $(n-k)$) qubits can be set simultaneously to $|1 \rangle$ in a qubit register of size $k$ (resp. $(n-k)$). 

Using the development (\ref{eq:devB}) in Eq. (\ref{eq:agp}), together with the definition of two sets of states (\ref{eq:agp-a}) and (\ref{eq:agp-bara}), we deduce the compact expression: 
\begin{eqnarray}
| K, n \rangle&=& \sum_{l=0}^K  \sqrt{\lambda^A_{l}} | l, k \rangle_A \otimes | K-l , n-k \rangle_{\bar A},
\label{eq:knsplit}
\end{eqnarray}
with 
\begin{eqnarray} 
{\lambda^A_{l}} &=&  \left[ C^l_K \right]^2  \frac{I^{k}_{l} (A)  I^{n-k}_{K-l} (\bar A) } {I^n_K} \equiv \frac{G^k_l(A) G^{n-k}_{K-l} (\bar A) }{G^n_k}, \label{eq:lambda} 
\end{eqnarray}
where, in the last expression, the $G^l_K$ coefficients are defined through $I^{k}_{l} = [l!]^2 G^k_l $.  The ${\lambda^A_{l}}$  are always positive and verifies the normalization conditions:
\begin{eqnarray}
\sum_{l=0}^K  {\lambda^A_{l}} = 1, \label{eq:normcond}
\end{eqnarray} 
so that these parameters can be interpreted as probabilities.  
Eq. (\ref{eq:knsplit}) is the exact Schmidt decomposition of the initial state when the total qubit register is split into two subspaces for any size $k$ of the subspace $A$. The constraints (\ref{eq:constraint}) are accounted for simply by assuming that $I^{k}_{l} (A) = 0$ (resp.  $I^{n-k}_{K-l} (\bar A)$) if $l > k$ (resp. $K-l > n-k$).  According to Eq. (\ref{eq:knsplit}), the reduced 
density has a simple diagonal structure:
\begin{eqnarray}
D_A = \sum_l \lambda^A_l D^A_l, \label{eq:decdensity}
\end{eqnarray}
where $D^A_l$ 
denotes the pure state density associated with the state 
$| l, k \rangle_A$. The corresponding entanglement entropy is given by: 
\begin{eqnarray}
S_{A} &=&  - \sum_{l = 0}^{K}  \lambda^{A}_l  \log_2 \lambda^{A}_l ,   \label{eq:entropdiag}
\end{eqnarray} 
from which we immediately deduce the upper bound $\log_2(K+1)$ for the reduced entropy. This upper bound 
might be further reduced depending on the $K$ values if we account for the constraints (\ref{eq:constraint}).
In the following, I will restrict to the case $0 \le k \le K \le n/2$. The upper bound becomes 
\begin{eqnarray}
S_A \le \log_2 (k+1). \label{eq:upper} 
\end{eqnarray}

Eq. (\ref{eq:knsplit}) is a generalization of well-known properties of spin systems with maximal spins. In this case, the 
set of coefficients $\{ \lambda^A_l \}_{l=1,K}$ can be obtained using standard techniques based on Clebsch-Gordan coefficients \cite{Coh86}. The decomposition 
(\ref{eq:knsplit}) can also be obtained for Dicke states using simple combinatoric arguments \cite{Mor18} (see also discussion below and some related discussions 
in Ref. \cite{Sto03}). 

\subsection{Technical details for the entropies evaluation}

For non-equal sets of $\{ x_i \}$ parameters, the entropies evaluations require 
the numerical calculations of the different coefficients given by Eq. (\ref{eq:lambda}). It could be shown that these coefficients 
can be accurately evaluated using some recurrence relations (see, for instance, \cite{San08,San09,Lac10,Hup11}).  Guided by the 
projection on particle number technique \cite{Rin80,Bla86}, I used an alternative numerical method based 
on the generating function described below. 

Let us assume that we consider a subset of parameters $\{ x_{i_1} , \cdots , x_{i_m} \}$ corresponding to a sub-system ${\cal S}$. This subset of parameters is associated with a certain subspace ${\cal S}$ of the total space. This subspace can be either the 
system $A$ itself, its complement $\bar A$, or the total space itself $(A+\bar A)$. We can then define the function:
\begin{eqnarray}
H_{\cal S}(\varphi) &=&  \prod_{i_k \in {\cal S}}  \left( 1 +|x_{i_k}|^2 e^{-i \varphi} \right) 
\end{eqnarray} 
for $\varphi \in \left[ 0, 2 \pi \right]$. Then, we have the property:
\begin{eqnarray}
G^m_l ({\cal S}) &=& \int_0^{2 \pi} \frac{d \varphi}{2 \pi}   e^{i l \varphi} H_{\cal S}(\varphi).
\end{eqnarray}
The integral can be accurately evaluated using the Fomenko technique \cite{Fom70}. Numerical results up to $m =50$ 
cannot be distinguished from the recurrence technique. For larger $m$ values, either an improved 
integration technique should be used, or the recurrence technique should be preferred. Note finally that, for a given separation 
of the total systems into two sub-systems, the three generating functions $H_{A}$, $H_{\bar A}$, and $H_{A+\bar A}$ should be 
used to compute the set of $\{ \lambda^A_l \}$ parameters through Eq. (\ref{eq:lambda}).           

\section{Illustrations of specific BTS states}
 \begin{figure}[htbp] 
\includegraphics[width=0.7\linewidth]{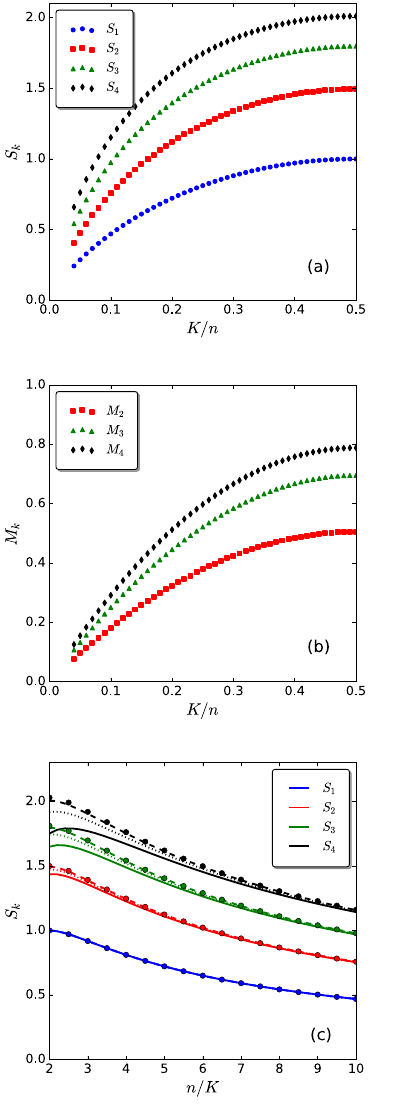}
    \caption{(a) $k$-qubit entropies for $k=1,\cdots,4$ of the state $| K, n \rangle$ obtained for the Dicke states for $n=100$ qubits with varying $4 < K \le n/2$ values. 
    (b) Corresponding mutual information defined in Eq. (\ref{eq:itermutual}) for $k=2, \cdots,4$. (c) Values of $S_{1,\cdots,4}$ for a state $| K, n \rangle$ with a fixed value of $K$ and increasing $n$ up to $n/K=10$. The solid line corresponds to $K=5$, dotted lines to $K=10$, while the dashed lines correspond to $K=40$. Note that all blue curves for $S_1$ are superimposed since, in this case, the one-qubit entropy is only a function of $K/n$. This figure also shows a set of curves displayed with filled circles corresponding to the binomial distribution with probability $P=K/n$. }
    \label{fig:Dicke1}
\end{figure}
    
\subsection{Dicke/Total spin BTS}

Dicke states correspond to the specific case where $x_i = {\rm cte}$ in Eq. (\ref{eq:agp}). For the sake of simplicity, I assume $x_i =1$ for all $i$ below.    
The entanglement entropies of these states have already been studied 
previously, for instance, in Ref. \cite{Mor18}. In particular, as further illustrated below, these states correspond to the maximally entangled states 
having the form given by Eq. (\ref{eq:agp}). The Schmidt decomposition of the Dicke states was given in Ref. \cite{Mor18} using simple combinatoric developments. 
Alternatively, using the generating functions discussed above, we can immediately show that:
\begin{eqnarray}
G^m_l ({\cal S}) &=& C^l_m , ~~{\rm for} ~l \le m \nonumber 
\end{eqnarray}
and $G^m_l ({\cal S}) = 0$ if $l > m $. This gives:
\begin{eqnarray}
 \lambda^{A}_l &=& \frac{C^{l}_k C^{K-l}_{n-k}  }{C^{K}_n}. 
\end{eqnarray}
 We recognize here the hypergeometrical (HG) probability distribution, which is properly normalized to $1$ thanks to the Chu–Vandermonde identity \cite{wikiChuVandermond}. A property that will be useful below is that the mean value and fluctuations of the Hamming weight, 
 denoted generically as $\mu_N (A)$ and $\sigma^2_N (A)$ in the subsystem $A$.  These two quantities are given by:
\begin{eqnarray}
\mu_N (A) = k  \frac{K}{n}, ~~\sigma^2_{N}(A) = k \left[ \frac{n - k}{n-1} \right] \frac{K}{n} \left( 1 - \frac{K}{n}\right). \label{eq:hgmean}
\end{eqnarray}

I illustrate here a few aspects of the Dicke state entanglement entropy. In Fig. \ref{fig:Dicke1}-a are shown the entanglement entropy $S_k$ for a subsystem 
of size $k=1$ to $k=4$ and for various state $| K, n \rangle$. Note that only the cases $K/n \le \sfrac{1}{2}$ are shown because the curve is symmetric 
with respect to this point. This could indeed be understood from the fact that the number of $1$ and the number of $0$ plays a symmetric role with respect to 
the vertical line $K/n = \sfrac{1}{2}$. Consequently, all $S_k$ with $k \le K$ have a bell shape for $0 \le K/n \le 1$ with a maximum exactly at $K/n=\sfrac{1}{2}$. In order 
to illustrate the entanglement between particles, I also give in Fig.  \ref{fig:Dicke1}-b, the corresponding mutual information $M_k$ that I define for $k \ge 2$ as:
  \begin{eqnarray}
M_k &=& S_{k-1} + S_{1} - S_{k}. \label{eq:itermutual}
\end{eqnarray}  
This quantity gives a measure of the entanglement of one particle selected within the set of $k$ particles. We observe that the mutual information increases but
this increase tends to be reduced as $k$ increases.         

For small numbers of qubits $k$, I observed that the $S_k$ rapidly become independent from the absolute value of $K$ itself, but only depends on the ratio $K/n$ 
provided that $K \le n/2$ is large enough and the number of sites $n$ itself is also large. A specific situation is the one-qubit entropy. In this case, we have
directly:
\begin{eqnarray}
S_1 &=& - \frac{K}{n} \log_2 \frac{K}{n} - \left( 1 - \frac{K}{n}\right) \log_2 \left(1 -\frac{K}{n}\right) \label{eq:ent1}
\end{eqnarray}
where we recognize the mean probability $P=K/n$ that the qubit is occupied. We see that $S_1$ is only dependent on the ratio $K/n$ and not on the specific value 
of $K$ or $n$ themselves.  In Fig. \ref{fig:Dicke1}-c, it is illustrated that this property is not true anymore for $S_k$ with $k > 1$, but the $S_k$ 
are close to each other for the different $K$ values. This stems from the fact that: 
 \begin{eqnarray}
 \frac{C^{K-l}_{n-k}  }{C^{K}_n}  &\longrightarrow &\left(\frac{K}{N}\right)^l \left(1 - \frac{K}{N}\right)^{k-l} ,
\end{eqnarray}
when both $K$ and $n$ are much larger than $l$ and $k$. 
 \begin{figure}[t] 
\includegraphics[width=0.8\linewidth]{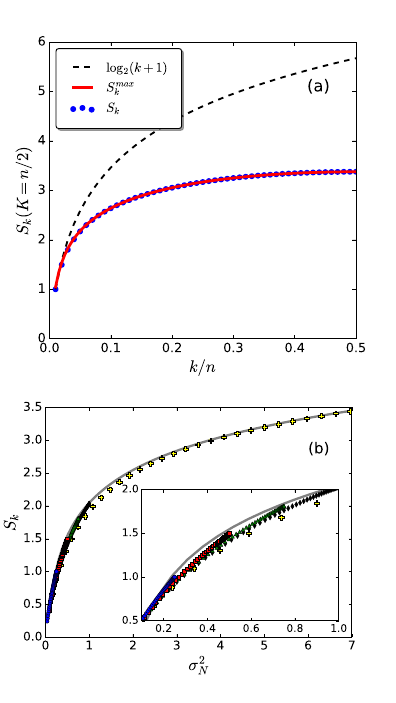}
    \caption{Panel a: Values of $S_k$ for the state $K=n/2$ with $n=100$ as a function of $k$. The black dashed line corresponds to the upper bound $\log_2 (k+1)$. The red solid line corresponds to the $S^{\rm max}_k$ values given by Eq. (\ref{eq:bound2}).
     Panel b: $S_{k}$ for $k=1,\cdots,4$ (with the same symbols convention as in Fig. \ref{fig:Dicke1}-a), for varying $K/n$ and $n=100$ as a function of the fluctuation of the Hamming weight $\sigma^2_N$ in the subspace $A$. The entropy is also shown for $k=K/2$ (yellow cross). 
     The grey solid line corresponds to the Gaussian approximation, Eq. (\ref{eq:GaussianLimit}). The inset is a zoom on the small $\sigma^2_N$ 
     values.  
}
    \label{fig:Dicke2}
\end{figure}
Accordingly, the corresponding coefficients $\lambda^A_l$ identify asymptotically with a binomial probability distribution that only depends on $P$. If we denote by $S^{\rm Bin}_k$ the entropy associated with a system of $k$ qubits where the equivalent binomial (bin) probability replaces the hypergeometric probability, we not only have $S^{\rm HG}_k \rightarrow S^{\rm bin}_k$, but also see in Fig. \ref{fig:Dicke1}-c that $S^{\rm bin}_k$ is an upper bound for the $S_k$ entropy.  This was checked numerically for all 
$(k,K)$ combinations up to $ n=100$.     

I conclude the discussion on the Dicke state entanglement properties by focusing on larger $k$ values in the system $A$. In Fig. \ref{fig:Dicke2}-a is shown the 
evolution of $S_k$ with increasing $k$ for the specific state $| K=n/2, n \rangle$, and $n=100$. For small $k$ values, the entropy is rather close to the upper bound 
given by Eq. (\ref{eq:upper}). As $k$ increases, we see a quenching of the entropy compared to this bound.  To understand this quenching, the entropies 
for selected $k$ are displayed in Fig. \ref{fig:Dicke2}-b as a function of the Hamming weight fluctuations given by Eq. (\ref{eq:hgmean}) for varying $K/n$ ratio and 
$m=100$. We see in this figure that all entropies are very close to each other, showing that the fluctuations are driving the amplitude of the different entropies. 
We finally see in this figure that the curves are rather also close to the asymptotic limit 
\begin{eqnarray}
\tilde S_k = \frac{1}{2} \log_2 (2 \pi e \sigma^2_N), \label{eq:GaussianLimit}
\end{eqnarray}
that is expected when both $K$ and
$n$ are large. Noteworthy, as shown in the inset, some deviations are observed for small $k$ values. 

Despite these small deviations, one can use the expression of the fluctuations given by  (\ref{eq:hgmean}) to write an approximate form of the entropy as (for $k \le K$):
\begin{eqnarray}
\tilde S_k &=&  \frac{1}{2} \log_2 \left[ 2 \pi e P (1-P)\right] +  \frac{1}{2} \log_2 k \left[ \frac{n - k}{n-1} \right].
\end{eqnarray}
Several considerations can be made from this simple expression. The first term only depends 
on the ratio $K/n$, while the subsystem size dependence is contained in the second term. 
The first term is maximal for $P=1/2$. Although I do not give here a firm mathematical proof, the present study 
suggests an upper bound for the Dicke state entanglement, denoted by ${S}^{\rm max}_{k}$ obtained by setting $P=1/2$
in the previous expression, and given by:  
\begin{eqnarray}
{S}^{\rm max}_{k} &=& \frac{1}{2} \log_2 \left[ \frac{\pi e}{2}\right] +   \frac{1}{2} \log_2 k\left[ \frac{n - k}{n-1} \right]. \label{eq:bound2} 
\end{eqnarray}
%
The constant term gives approximately $c_0 =  \frac{1}{2} \log_2 \left[ \frac{\pi e}{2}\right]  \simeq 1.047$. We see that 
we have respectively $S^{\rm max}_1 = 1.047$,  $S^{\rm max}_2 = 1.55$,  $S^{\rm max}_3 = 1.84$,  and $S^{\rm max}_4 = 2.05$ that are just above 
the highest values of these entropies displayed in Fig.  \ref{fig:Dicke1}-c for $n/K =2$. We also see that ${S}^{\rm max}_{k} $ 
follows very closely the numerical values reported in Fig. \ref{fig:Dicke2}-a, and properly accounts for the reduction compared to the limit given by
(\ref{eq:upper}).  Note that, if we neglect the factor $(n - k)/(n-1)$ in Eq. (\ref{eq:bound2}), we get a slightly larger value ${S}^{\rm max}_{k} \sim \frac{1}{2} \log_2 k+1$, that was obtained in Ref. \cite{Sto03}. However, the correction factor in Eq. (\ref{eq:bound2}) is important at large $k$ to reproduce the numerical values.

 \begin{figure}[htbp] 
\includegraphics[width=0.8\linewidth]{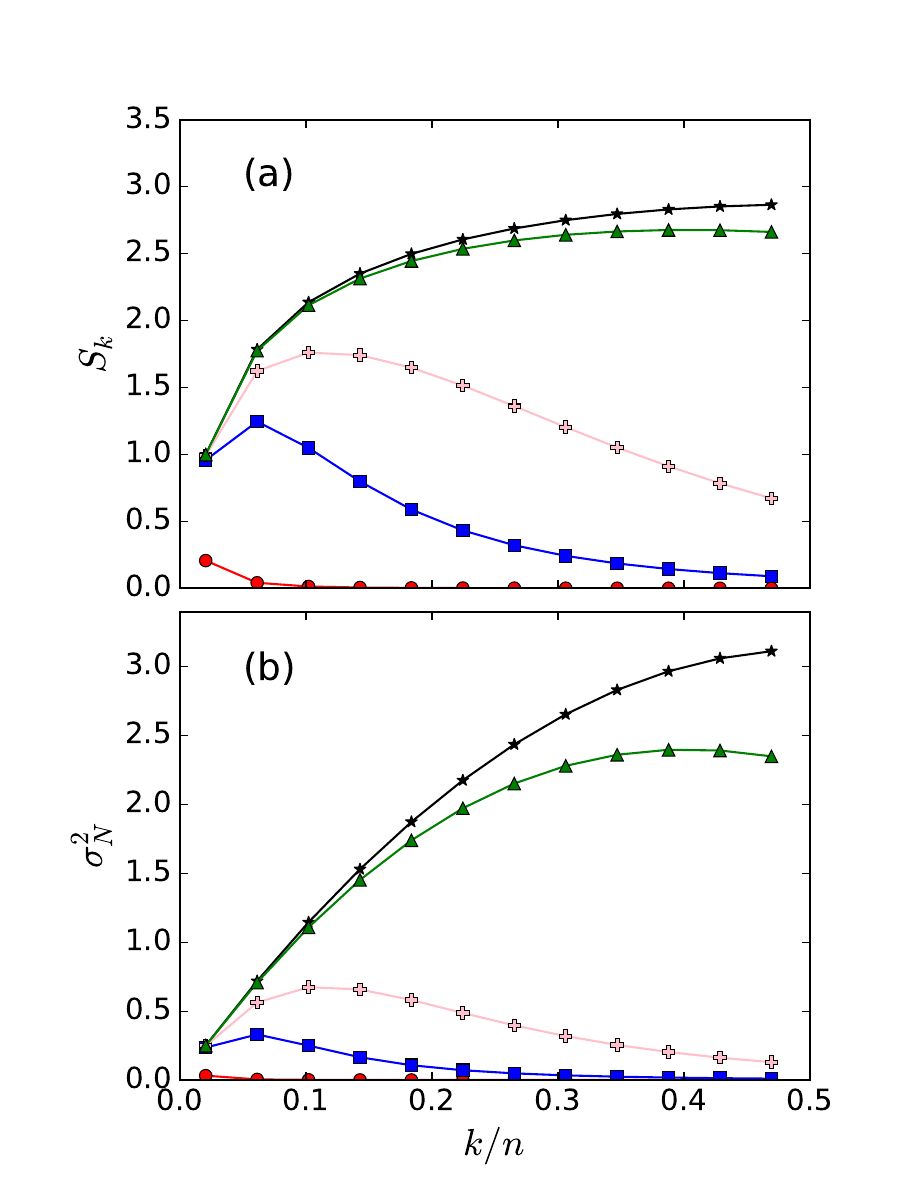} 
    \caption{Illustration of the $k$-qubit/pairs entropy using BTS state inspired by the projected BCS theory (see text).
(a)  The entropy $S_k$ where $k$ is the number of "qubits/pairs" within a certain energy window defined by an increasing $E_{\rm cut}$  is shown as a function 
of $k/n$. (b) fluctuations of particle number in the same energy window. The black solid line corresponds to the reference "maximally entangled" Dicke state, e.g. with $|x_i|^2=1$ 
for all $i$. Different values of the parameter $\Delta / de$ are shown: $\Delta / de = 1$ 
(red circles), $5$ (blue squares), $10$ (pink cross) and $50$ (green triangle). In all calculations, I used $n_{\rm st} = 20$ ($n=49$) and assumed $K = n_{\rm st}$.   
}
    \label{fig:BCS-BTS}
\end{figure}

 \begin{figure}[htbp] 
\includegraphics[width=0.9\linewidth]{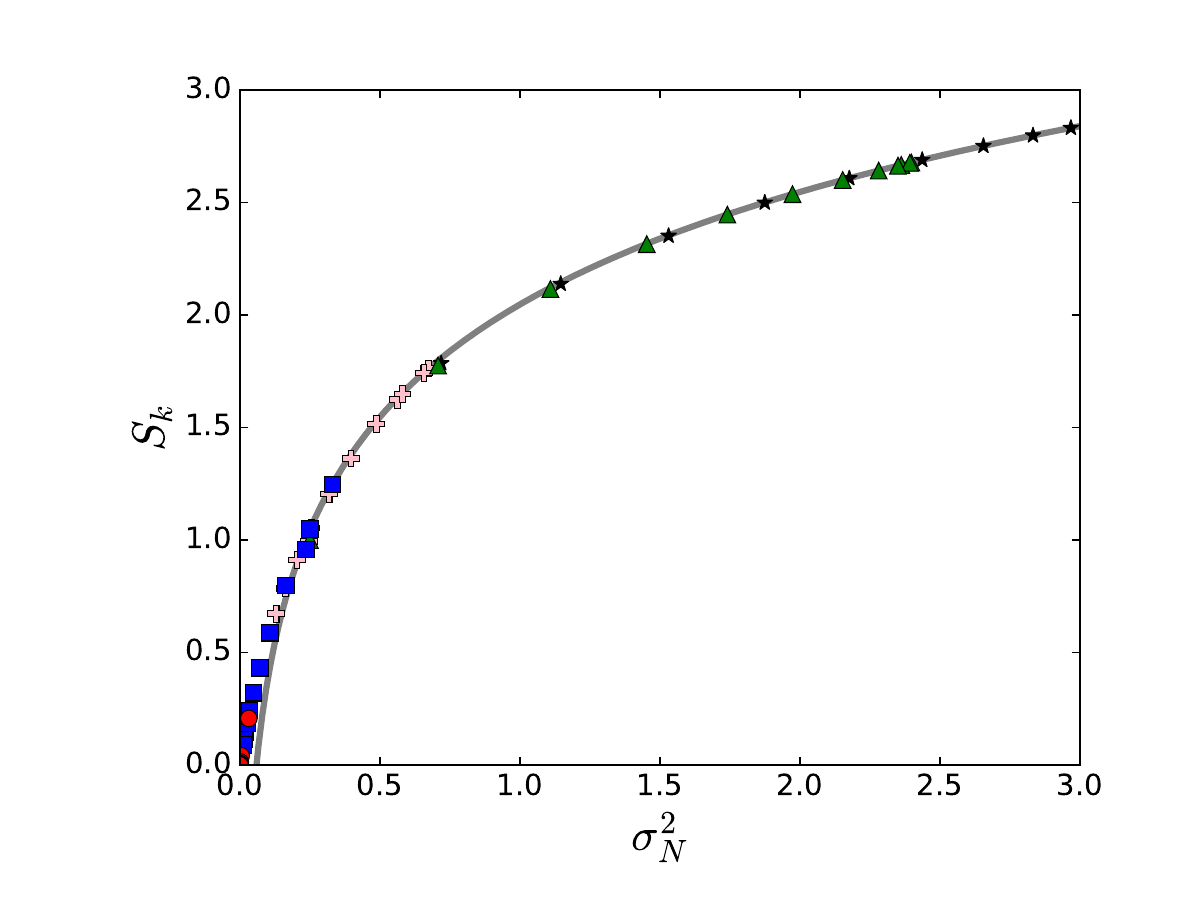} 
    \caption{Same as in Fig. \ref{fig:BCS-BTS} except that now the entropy is shown as a function of the qubit/pair number 
    fluctuations in the energy window. The conventions are the same as in Fig. \ref{fig:BCS-BTS} except that the lines connecting symbols are omitted for clarity.
    As in Fig. \ref{fig:Dicke2}, the gray solid line corresponds to the Gaussian limit given by Eq. (\ref{eq:GaussianLimit}).
}
    \label{fig:skvssigma}
\end{figure}

\subsection{Projected BCS state}

The projected BCS states are a second illustration of BTS states that motivated the present work. These states are used for small superfluid systems. 
Starting from their second quantized form given by Eq. (\ref{eq:bcs}), and using the pair encoding scheme \cite{Kha23}, BCS states can be expressed in terms of Pauli matrices as:
 \begin{eqnarray}
| \Phi (\{ x_i \}) \rangle &=& {\cal N}_m \prod_{i=1}^{m} \left( 1 + x_i \sigma^+_i \right) | 0 \rangle^{\otimes m} . \label{eq:bcsencoding} 
\end{eqnarray}
The $U(1)$ symmetry breaking associated with the fact that BCS states are not eigenstates of particle number transforms here into a mixing of states 
with different Hamming weights. Then, projected BCS is equivalent to projecting onto a specific Hamming weight equal to $K$, and after proper normalization, 
identify with the BTS state given by Eq. (\ref{eq:agp}).  It is interesting to mention that, although projected states will have non-trivial entanglement between pairs
of particles, the BCS state given by (\ref{eq:bcsencoding}), being written as simple unary operations on individual qubits, leads to zero mutual
information between pairs.      
  
To make contact with the standard physicist's approach, I will consider below a set of states described by their single-particle energies $\{\varepsilon_i \}_{i=1,n}$
and assume that we have already solved BCS-like equations leading to a set of coefficients $(u_i,v_i)$.   Having in mind the constant interaction pairing model, I parameterize these coefficients with two free parameters $(\lambda, \Delta)$:
\begin{eqnarray}
|u_i|^2 = \frac{1}{2}\left( 1 + \frac{\varepsilon_i - \lambda}{\sqrt{(\varepsilon_i - \lambda)^2 + \Delta^2}} \right), \nonumber \\
|v_i|^2 = \frac{1}{2}\left( 1 - \frac{\varepsilon_i - \lambda}{\sqrt{(\varepsilon_i - \lambda)^2 + \Delta^2}} \right). \nonumber 
\end{eqnarray}   
The $x_i$ parameters in Eq. (\ref{eq:bcsencoding}) can then be obtained from:
\begin{eqnarray}
|x_i|^2 &=& \frac{|v_i|^2}{|u_i|^2} = \frac{1 - \frac{\varepsilon_i - \lambda}{\sqrt{(\varepsilon_i - \lambda)^2 + \Delta^2}} }{ 1 + \frac{\varepsilon_i - \lambda}{\sqrt{(\varepsilon_i - \lambda)^2 + \Delta^2}} } . \label{eq:bcsxi}
\end{eqnarray} 
Physically, $\lambda$ and  $\Delta$ represent the Fermi energy and pairing gap, respectively. 
In the following, I assume that all $x_i$ are positive real numbers. Noteworthy,  in the limit $\Delta \rightarrow + \infty$, all $|x_i|^2$ tend to $1$, and we recover the Dicke state limit. 

As a numerical illustration, I  consider below a set of $n = 2 n_{\rm st}+1$ equidistant levels with energies $\varepsilon_i$ distributed in $[-E_{\rm max}, E_{\rm max}]$, and $\lambda=0$. I assume that $\Delta$ is a free parameter. Calculations below are made assuming $n=49$. We then have $E_{\rm max} = n_{\rm st} de $, where $de$ defines the level spacing. Single-particle energies are given by  $\varepsilon_i = de (i - 1 - n_{\rm st})$ with $i=1,\cdots,n$. All energies will be given in $de$ units. 
I then assign to each site a $x_i$ coefficient given by Eq. (\ref{eq:bcsxi}) obtained by fixing the gap value $\Delta$. As before, 
the different qubits are separated into two subspaces $A$ and $\bar A$ and, for different splitting into two subspaces. I compute the reduced entropy of the subspace $A$ with varying numbers of sites included 
in $A$ and/or varying values of $\Delta$. In the illustration below, the subspace $A$ is built up as follows.  $A$ corresponds to the subset of states 
respecting the condition $|\varepsilon_i| \le E_{\rm cut}$, where $E_{\rm cut}$ is a cutoff energy, that we assume 
equal to $E_{\rm cut}/de = 1/2 + k_c$.  The subspace size is then gradually increased by varying $k_c$ with $k_c = 0, \cdots , n_c$ and $n_c = n_{\rm st}/2$ (here $n_c = 10$). For a given $k_c$,  the number of qubits/orbitals inside the energy window verifies $k= 1+ 2 k_c$. 

In Fig.  \ref{fig:BCS-BTS}-a,
the $k$-qubit entropy is displayed as a function of $k$, i.e., for increasing energy window size and for different values of $\Delta/de$. 
Panel b displays the fluctuations in the subspace $A$ of the pair number/Hamming weight of the projected BCS state.  In Fig.  \ref{fig:BCS-BTS}, we see again that fluctuations in the subsystem $A$
and entropies are strongly correlated. As expected, as $\Delta$ increases, the entanglement entropy tends towards the Dicke state limits, corresponding to 
the upper limit of the projected BCS-type states. The strong correlations between entanglement and fluctuations are further evidenced in Fig. \ref{fig:skvssigma}, where we 
recognize a similar behavior as we already observed in Fig. \ref{fig:Dicke2}-b. We also see that, as soon as the fluctuations exceed a certain threshold, the correlations between $S_k$ and $\sigma^2_N$ closely follow the Gaussian asymptotic limits. Interestingly, this threshold that is around $\sqrt{\sigma^2_N} \simeq 0.5$ is rather low since it corresponds to a fluctuation of less than one particle unit. The validity of the Gaussian limit when considering part of
a small superfluid, even for rather small systems, was already pointed out in Ref. \cite{Lac20}.  
Below this threshold in fluctuations, the Eq. (\ref{eq:GaussianLimit}) slightly underestimates the entropies uncovering the non-Gaussian nature of the
fluctuations for very small subsystems.       

\section{Discussion on states decomposed on BTS states}

In the present work, I concentrate on the specific entanglement entropy of a single BTS state $| K, n \rangle$.
States like Dicke states provide a convenient basis when a system is described on a set of degenerated two-level systems labeled by $i=1,n$ and when the problem 
is invariant with respect to any permutation between indices of $(i_1, \cdots, i_n)$ labeling the 2-levels. An example of such a physical situation 
is the Lipkin model \cite{Lip65,Mes65,Gli65}.  
Here, I discuss a situation where the evolution of a permutation invariant system is considered. Its wave function can be decomposed at all times on the Dicke states as:
\begin{eqnarray}
| \Psi(t) \rangle &=& \sum_K c_K(t) | K, n \rangle .  \label{eq:timevol}
\end{eqnarray}
The block structure that has been made explicit above for each of the $| K, n \rangle$ state also exists for a state written 
as a linear combination of BTS states.  
To see this, I again split into two parts $(A, \bar A)$ containing respectively $k$ and $(n-k)$ 
qubits/orbitals. I also rewrite Eq. (\ref{eq:knsplit}) in a more compact form as:  
\begin{eqnarray}
| K, n \rangle&=& \sum_{l=0}^K  g^K_l | l \rangle_A \otimes | K-l \rangle_{\bar A}, \nonumber 
\end{eqnarray}   
with $g^K_l = \sqrt{\lambda^A_{l} (K)}$, $| l, k \rangle_A \equiv | l \rangle_A $, and $| K-l , n-k \rangle_{\bar A} \equiv | K-l \rangle_{\bar A}$.
Conditions given by Eq. (\ref{eq:constraint}) are accounted for simply by assuming that the $g^K_l = 0$ if these conditions are not fulfilled.  
With this, we immediately see that the reduced density matrix of the subsystem $A$ is given by:
\begin{eqnarray}
D_A(t) &=& \sum_{j,l=0}^k | j \rangle_A \left[ {\cal D}^A_{jl} \right] {}_{A} \langle l |  . 
\end{eqnarray} 
with the matrix elements 
\begin{eqnarray}
{\cal D}^A_{jl}(t) &=& \sum_{K,L=0}^{n}  c_K(t) c^*_L(t) g^K_j  g^J_l \delta_{K-j,L-l}. 
\end{eqnarray} 
These matrix elements can be computed for any $k$ using the expression (\ref{eq:upper}). This expression clearly underlines that 
only the coupling between permutation invariant states of the sub-system $A$ is relevant to discuss the entanglement entropy 
and that the subsystem entropy is also bound by Eq. (\ref{eq:upper}), as expected from the symmetry of the problem. It also gives
a practical way to calculate the entropy itself for any binary partition of the total system.  Previous studies of the entanglement 
entropy for systems like the Lipkin model \cite{Fab22,Mom23} or neutrino \cite{Lac22} were mostly restricted to $k=1$ or $k=2$ and directly used the naive binary decomposition 
of the subsystem as schematically illustrated in Fig. \ref{fig1:bts}. In Ref. \cite{Mom23}, the study of the 4-qubit entanglement entropy was made following the same strategy. 
This brute-force approach, which does not take advantage of the permutation symmetry 
rapidly becomes cumbersome and leads, for a given $k$, to the diagonalization of a matrix of size $2^k$. With the present method, the matrix 
$D_A(t)$ has a much lower dimension equal to $(k+1)$.   

Contrary to the case of a single Dicke state where the entropy is bounded by Eq. (\ref{eq:bound2}), during the evolution of a system decomposed 
as in Eq. (\ref{eq:timevol}), larger entropies can be reached due to the additional mixing between Dicke states. In Ref. \cite{Lac22}, the absolute 
upper bound given by Eq. (\ref{eq:upper}) was reached by coupling two different sets of two-level systems, called in this context neutrino beams, 
and allowing the simultaneous mixing of Dick states having both odd and even Hamming weight during the evolution.

\section{Summary and discussion}
  
The binary tree structure of certain quantum ans\"atz allows us to derive compact forms of their entanglement entropies
when the system is separated into two subsystems. The exact Schmidt decomposition of k-qubit 
can be obtained and eventually estimated at low numerical cost. With this, I study some quantum information properties of specific BTS states.
The Dicke state's scaling properties with fluctuations and upper bound of the entropies are analyzed. Since these states correspond 
to maximally entangled states having the BTS structure with fixed Hamming weight, the upper bound in entropy also holds 
for any BTS of such kind.

A second study is conducted on the entanglement properties of pairs in the projected BCS states. It is observed again that   
this entanglement is strongly correlated to the Hamming weight/pair number fluctuations and, in most situations very 
well described by the Gaussian entropy limit.  This second class of states is standardly used in nuclear physics to describe 
in an effective way static and dynamical properties when the system wave-function is assumed to be a single quasi-particle vacuum. 
The properties derived here, therefore, should apply to this case, for instance, when cutting a system into two sub-parts.    

I finally show that the technique developed here to numerically estimate the entanglement properties of BTS can also be useful 
to estimate similar entropies in systems whose wave functions decompose on a set of BTS states. 

\noindent {\it Acknowledgments. }-- The author is thankful to A. B. Balantekin, Amol V. Patwardhan, and Pooja Siwach 
for discussions at the early stage of the project especially on neutrino physics aspects. 
This project has received financial support from the CNRS through
the AIQI-IN2P3 project.  This work is part of 
HQI initiative (www.hqi.fr) and is supported by France 2030 under the French 
National Research Agency award number "ANR-22-PNQC-0002".

\end{document}